\documentclass[pre,
 reprint,
superscriptaddress,
 amsmath,amssymb,
 aps,
]{revtex4-2}
\usepackage{rotating}
\usepackage{multirow}
\usepackage{comment}
\usepackage{graphicx}
\usepackage{dcolumn}
\usepackage{bm}
\usepackage{ulem}
\usepackage[dvipsnames]{xcolor}

\newcommand\p[2]{\ensuremath{\frac{\partial #1}{\partial #2}}}

\def\half{{\textstyle \frac{1}{2}}}
\def\third{{\textstyle \frac{1}{3}}}
\def\fourth{{\textstyle \frac{1}{4}}}

\def\p#1#2{{\ensuremath{\frac{\partial #1}{\partial #2}}}}

\def \da {\ensuremath{\delta \hat{\alpha}}}

\def \thetamax {\ensuremath{{\theta_{\max}}}}
\def \thetaosc {\ensuremath{{\theta_{\mathrm{osc}}}}}
\def \tosc {\ensuremath{{t_{\mathrm{osc}}}}}
\def \phiosc {\ensuremath{{\phi_{\mathrm{osc}}}}}
\newcommand{\mi}{\mathrm{i}}

\begin{document}

\preprint{APS/123-QED}




\title{How do imperfections cause asymmetry in elastic snap-through?}

\author{Andrea Giudici}\thanks{These authors contributed equally.}
\affiliation{Mathematical Institute, University of Oxford, Woodstock Rd, Oxford, OX2 6GG, UK}

\author{Weicheng Huang}\thanks{These authors contributed equally.}
\affiliation{School of Engineering, Newcastle University, Newcastle upon Tyne NE1 7RU, UK}

\author{Qiong Wang}
 \affiliation{Department of Mechanical Science and Engineering, University of Illinois at Urbana-Champaign, Urbana, IL 61801}

 \author{Yuzhe Wang}
\affiliation{Singapore Institute of Manufacturing Technology, Agency for Science, Technology and Research, Singapore, 138634, Singapore}

\author{Mingchao Liu}
\affiliation{Department of Mechanical Engineering, University of Birmingham, Birmingham, B15 2TT, UK}
\affiliation{School of Mechanical and Aerospace Engineering, Nanyang Technological University, Singapore 639798, Singapore}

\author{Sameh Tawfick}%
\affiliation{Department of Mechanical Science and Engineering, University of Illinois at Urbana-Champaign, Urbana, IL 61801}

\author{Dominic Vella}%
 \email{dominic.vella@maths.ox.ac.uk}
\affiliation{Mathematical Institute, University of Oxford, Woodstock Rd, Oxford, OX2 6GG, UK}

\date{\today}
\date{\today}

\begin{abstract}

A symmetrically-buckled arch whose boundaries are clamped at an angle has two stable equilibria: an inverted and a natural state. When the distance between the clamps is increased (i.e.~the confinement is decreased) the system snaps from the inverted to the natural state. Depending on the rate at which the  confinement is decreased (`unloading'), the  symmetry of the system during  snap-through may change: slow unloading results in snap-through occurring asymmetrically, while fast unloading results in a  symmetric snap-through. It has recently been shown [Wang \emph{et al.}, \emph{Phys. Rev. Lett.} \textbf{132}, 267201 (2024)] that the transient asymmetry at slow unloading rates is the result of the amplification of small asymmetric precursor oscillations (shape perturbations) introduced dynamically to the system, even when the system itself is perfectly symmetric. In reality, however, imperfections, such as small asymmetries in the boundary conditions, are present too. Using numerical simulations and a simple toy model, we discuss the relative importance of intrinsic imperfections and initial asymmetric shape perturbations in determining the transient asymmetry observed. We show that, for small initial perturbations, the magnitude of the asymmetry grows in proportion to the size of the intrinsic imperfection but that, when initial shape perturbations are large, intrinsic imperfections are unimportant --- the asymmetry of the system is dominated by the transient amplification of the initial asymmetric shape perturbations. We also show that the dominant origin of asymmetry changes the way that asymmetry grows dynamically. Our results may guide engineering and design of snapping beams used to control insect-sized jumping robots.

\end{abstract}

\maketitle

\section{Introduction}

Elastic snap-through occurs when an elastic structure initially has two stable equilibria but one of these equilibria is lost (or becomes unstable) forcing the system to suddenly jump to the other remaining stable state. Although the initial part of the dynamics of snap-through may be slow \cite{Gomez2017, Sano2019}, the bulk of the dynamics is typically fast. This property is exploited by nature to create fast motion. For example,  Venus flytraps use snap-though to catch their prey \cite{Forterre2005} while click beetles use snap-through to right themselves after landing on their back \cite{Bolmin2021}.

In engineering too, snap-though instabilities can be harnessed to obtain fast motion. Simple examples include the popular jumping toys \cite{Isenberg1987,Ucke2009,Pandey2014} that exploit the fast inversion of a spherical cap hitting the ground to propel the cap into the air. A similar principle, with a beam undergoing a fast change of configuration and hitting the has been used to design insect-sized robots capable of large and repeated jumps \cite{Wang2023}. Much attention has thus been dedicated to designing a snap through mechanism that allows control and optimization of jumps \cite{tong2024inverse}. 

Controlling the behaviour of engineering systems exploiting snap-thorough requires us to understand the transient (a-)symmetry of the observed shapes. For example, consider the dynamical behaviour of an arch whose ends are clamped at an angle $\alpha$ to the horizontal. When the two ends of the system are sufficiently close, the arch has two stable equilibria: a natural state (resembling a V) and an inverted state (resembling an W). If the system is in the inverted (W) state and the two ends are released (unloaded) at a constant rate, eventually the system snaps to the other equilibrium (V). Although a linear stability analysis predicts that the transition occurs via asymmetric modes, it was recently shown that the symmetry of the transient shape during snap through depends on the unloading rate: slow unloading rates exhibit large asymmetry while fast unloading rates maintain symmetry \cite{qiong2024transient}. 

In the case of precisely identical clamp angles (i.e.~perfect symmetry), \citet{qiong2024transient} showed that the large observed asymmetry during snap-through is a result of the amplification of an initial asymmetric shape perturbation: small asymmetric precursor oscillations are inevitably introduced by the unloading prior to snap-through and these are then amplified during snap-through if the unloading is sufficiently slow \cite{qiong2024transient}. We shall refer to such oscillations \emph{before} snap-through as  `precursor oscillations'.

In most real systems, however, obtaining perfectly symmetric conditions is difficult, if not impossible; imperfections will always exist, leading to the question: how do imperfections affect the transient asymmetry of a snapping beam? In particular,  when do intrinsic imperfections dominate the dynamics and when is it instead precursor oscillations that dominate? 
Answering this question is fundamental to control snap-through in applications and might be required, for example, to control the jump direction of jumping robots \cite{Wang2023}.

In this article, we use numerical simulations and a simple toy model \cite{qiong2024transient} to answer the aforementioned questions by studying how imperfections --- in the form of a small asymmetry in the clamping angles --- affect the symmetry of  snap-through dynamics in a clamped arch. We begin by studying how the system behaves with only intrinsic imperfections (i.e.~in the absence of precursor oscillations). In this case, we show that the size of the transient asymmetry is proportional to the initial size of the imperfection and is largest for slow loading rates. We then investigate how a combination of an intrinsic imperfection and an initial shape perturbation (precursor oscillation) determines the behaviour of the system. We show that if the precursor oscillations introduced by the perturbation are small, the system remains dominated by the imperfection. When the precursor oscillations from the initial perturbation are relatively large, however, the system behaves essentially as in the perfectly symmetric case \cite{qiong2024transient}.  Finally, we discuss how the experimental measurements presented in \cite{qiong2024transient} are explained by a combination of imperfections and perturbations: slow unloading rates tend to be dominated by imperfections while fast unloading rates are heavily influenced by precursor oscillations.

\section{Theory and numerical simulations }

\begin{figure*}[t]
\includegraphics[width=\textwidth]{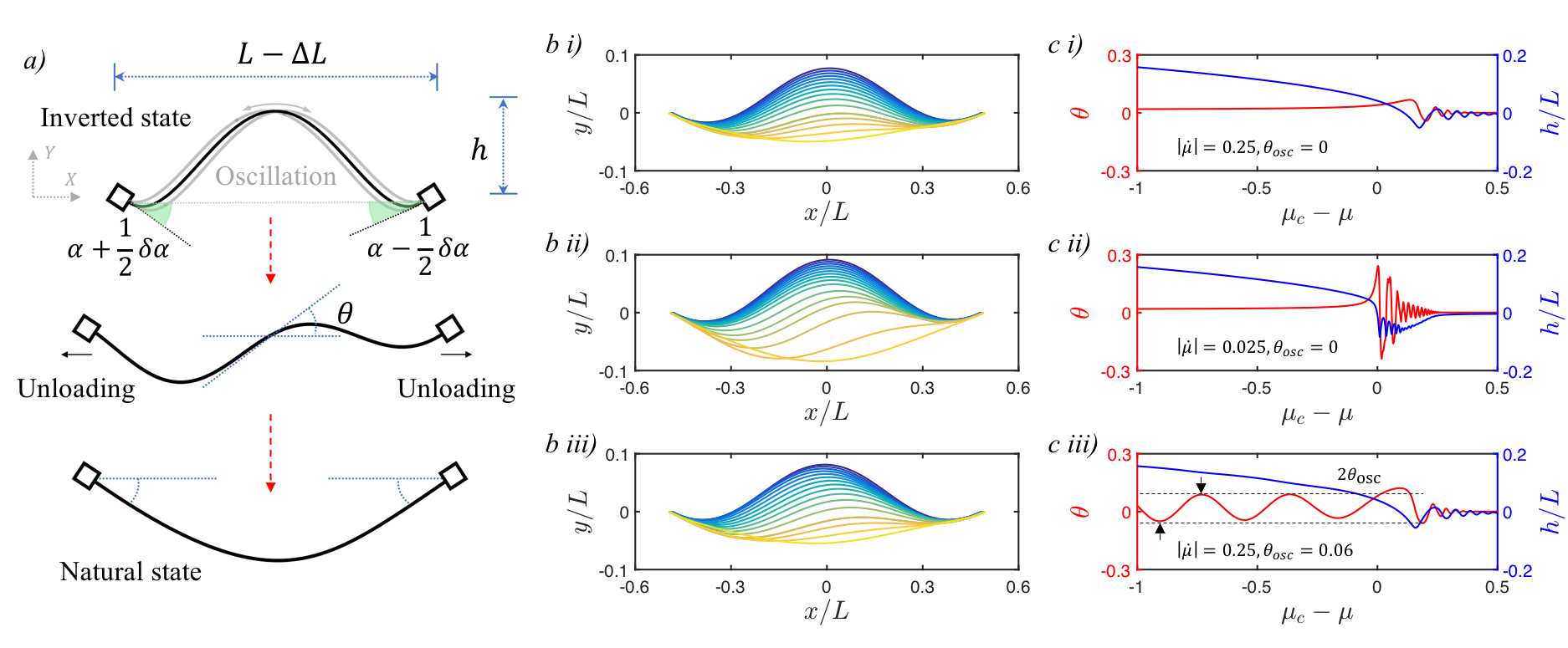}
\caption{Observed transient asymmetries during the snap-through of an arch. a) sketch of the beam with asymmetric clamps in the inverted (top) and natural state (bottom). The mid-point angle is $\theta$ (radians), the asymmetry during snap-through. b) Numerical simulations of the shape evolution of an arch with a small imperfection of the clamps angle,  $\da=\delta\alpha/\alpha=0.1$, for different unloading rates, as described in text in the plots of panels (c). For fast rates (top), the shape remains symmetric at all stages. In the middle image, slower unloading leads to significant growth of the intrinsic asymmetry during snap-through. In the bottom image, the initial condition introduces small asymmetric precursor oscillations which are amplified during snap-through. c) Shows the  evolution of the central angle, $\theta$, and central height, $h$, as $\dot{\mu}$ changes, highlighting that slower rates lead to a larger mid-point angles and, thus, greater asymmetry.}
\label{fig:overviewPlot} 
\end{figure*}

\subsection{Model of a snapping arch}

The elastic arch used experimentally is an elastic strip of  length $L$, thickness $H$ and density $\rho$. Because of the imperfection in clamping, we assume that it is clamped such that its two ends make an angle with the horizontal of $\alpha + \delta \alpha/2$ and $\alpha- \delta \alpha/2$ on the left and right, respectively. (The beam is thus clamped symmetrically when $\delta \alpha=0$.) The arch is made of isotropic material with Young's modulus $E$ (so that its bending stiffness $B=E H^3/12$) and is compressed horizontally by a distance $\Delta L$ --- the imposed `end-shortening'. Its mid-point height is then $h$. The system is shown in figure \ref{fig:overviewPlot}a.

\begin{figure*}[t!]
\includegraphics[width=\textwidth]{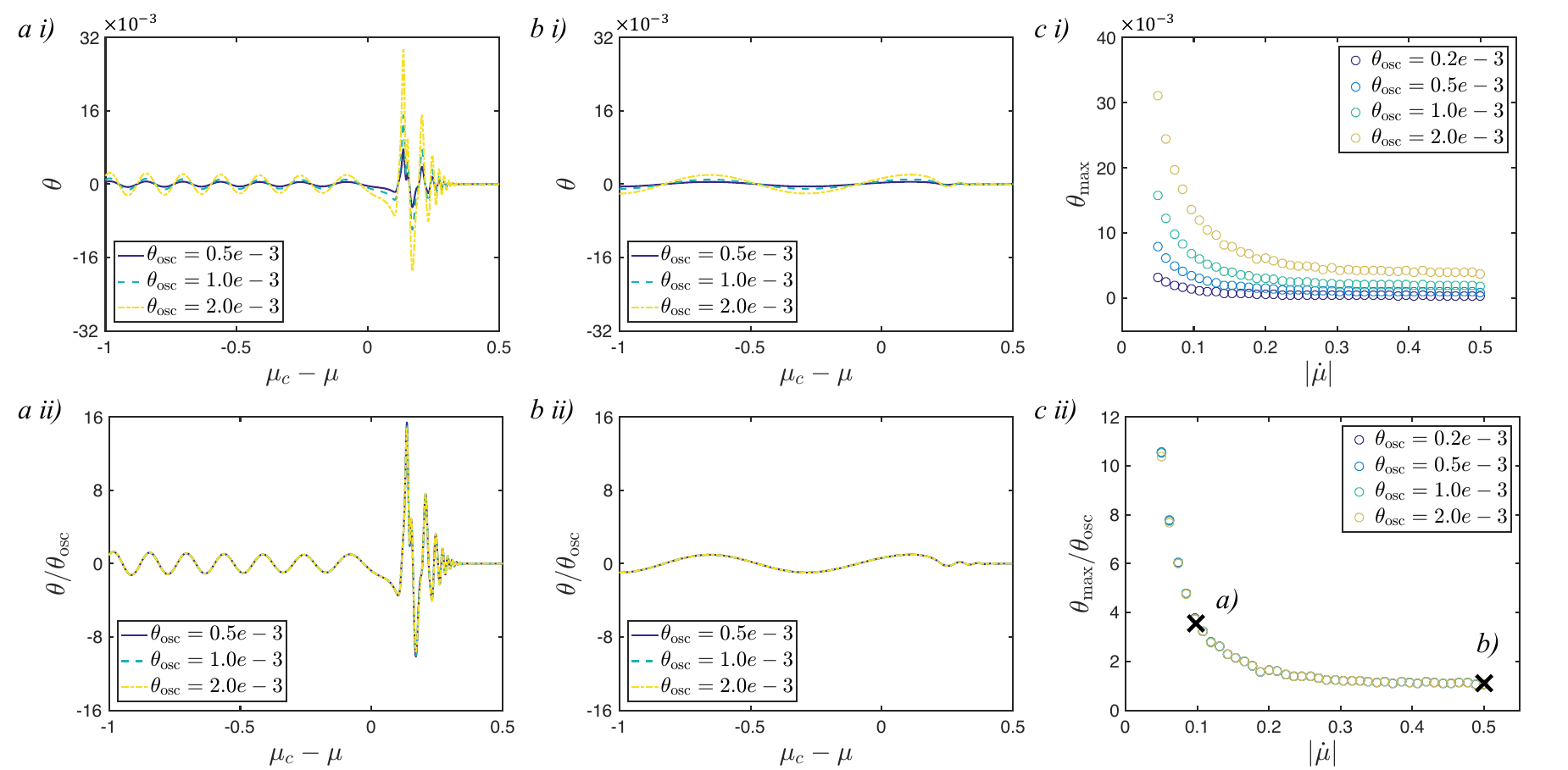}
\caption{
Results of numerical simulations of the snap-through of an arch with perfectly symmetric boundary conditions, $\da=0$, as studied in ref.~\cite{qiong2024transient}. Asymmetry is shown via the evolution of the mid-point angle, $\theta$, with distance to the first bifurcation  ($\mu_c=1/4$) for different sized asymmetries in the initial condition, $\thetaosc$, and for slow (a) and fast (b) loading rates. In (a-c), the top row (i) shows raw numerical results while the bottom row (ii) shows that rescaling by $\thetaosc$ leads to a collapse of data and hence that the amplification, ${\cal A}=\theta/\thetaosc,$ is the appropriate measure of asymmetry in this case. (a) For slow unloading,   $|\dot{\mu}|=0.1$, $\theta$ increases dramatically in the vicinity of the bifurcation point ($\mu-\mu_{c}=0$). (b) For faster unloading rate, $|\dot{\mu}|=0.5$, $\theta$ does not noticeably grow. (c) The maximal mid-point angle observed during snap-through, $\thetamax$,  shows that the observed asymmetry increases significantly as $|\dot{\mu}|$ decreases. 
}
\label{fig:oscillationPlot} 
\end{figure*}

For sufficiently large end-shortening, the arch is bistable. However, as the system end-shortening is decreased, the inverted state ceases to exist and the arch snaps to the natural state. To characterize what makes a given $\Delta L$ large or small, previous authors \cite{qiong2024transient,Liu2021}  have noted that the typical angle induced by the compression, $(\Delta L/L)^{1/2}$ may be compared to the imposed angle, $\alpha$. These authors \cite{qiong2024transient,Liu2021} have therefore introduced the dimensionless parameter
\begin{equation}
    \mu= \frac{1} {\alpha^2} \frac{\Delta L}{  L}.
\end{equation} For convenience, the parameter $\mu$ is referred to in this paper as the `end-shortening' --- the rescaling by $\alpha^2$ being understood. In \cite{qiong2024transient}, it was shown that at $\mu=1/4:=\mu_c$, the symmetrically clamped arch becomes unstable to asymmetric perturbations, while at $\mu\approx0.247:=\mu_\ast$, the inverted equilibrium is lost in a saddle-node bifurcation.

In this paper, another important dimensionless parameter emerges: the relative size of imperfection to the intrinsic angle, which we denote 
\begin{equation}
    \da =\frac{\delta \alpha}{\alpha}.
\end{equation}
Note that when ($\da \neq 0$) the midpoint angle is not zero at equilibrium ($\theta\neq 0$); the imperfection in the clamps leads to a small asymmetry in the shape. 

To observe  snap-through and study its symmetry, we begin with a sufficiently large compression ($\mu>0.3$) and place our arch in the inverted state. We then decrease the end-shortening at a constant rate $\dot{\mu}<0$, with
\begin{equation}
    \dot{\mu}=\frac{\dot{\Delta L}\,t^*}{\alpha^2 L},
\end{equation}
 and dots representing differentiation with respect to time $t$; here $t_*=0.108 \cdot L^2 \sqrt{{\rho bh}/{B}}$ 
is the typical timescale of oscillations in the natural state at $\mu=1/4$ \cite{qiong2024transient}.

As the end-shortening is decreased, the system eventually snaps to the natural state. The symmetry of the transient shape during snap-through depends on the rate and the size of initial perturbations introduced in the system. 
In the experimental setup discussed in \cite{qiong2024transient}, these precursor oscillations were introduced as a byproduct of the experimental procedure: the unloading was done with a linear motor connected to only one of the clamps meaning that, as the experiment is initiated, and the clamp is accelerated (very quickly) to its final velocity, an asymmetric impulse is given to the beam generating small asymmetric oscillations. We measure the symmetry of the transient shape of the system via the angle the arch makes with the horizontal at its midpoint, $\theta$, where perfectly symmetric shapes maintain $\theta=0$ throughout.  We denote the amplitude of the precursor oscillations by $\theta_{\mathrm{osc}}$.

\subsection{Numerical model and solution}

Our numerical simulations use the discrete elastic rods algorithm \cite{bergou2008discrete,bergou2010discrete, huang2019newmark}, following the methods used to model a snapping arch  described in \cite{qiong2024transient}. To control the role of the precursor oscillations, $\thetaosc$, we use a symmetrical loading protocol (moving both ends together --- each at half the total speed --- at the same time); in this way no precursor oscillations are excited directly by loading. Instead, precursor oscillations are introduced by applying a transverse body force $F$ for a short time $\Delta t$ (i.e.~the whole beam is pushed to the left or right briefly). In this way, the amplitude of the precursor oscillations can  be controlled easily by varying the size of $F$. This approach is different to the experiments (in which  only one end of the arch is attached to a motorized linear stage and precursor oscillations are excited by this loading asymmetry), but is more convenient for numerical simulations. 
Panels (b) and (c) of Fig.~\ref{fig:overviewPlot} show the results of three simulations for the snap-through of an imperfect arch with $\da=0.1$ but different end-shortening rates, $\dot{\mu}$,  and size of the precursor oscillations, $\thetaosc$. Comparing the top and middle rows shows the influence that the loading rate has: with no precursor oscillations, the shape asymmetry remains small for fast unloading rates (fig.~\ref{fig:overviewPlot}c(i)), but becomes significant at slow rates (fig.~\ref{fig:overviewPlot}c(ii)); this change in the degree of asymmetry is also clear in the snap-shots of the shape shown in fig.~\ref{fig:overviewPlot}b(i)-(ii). Even at large rates, significant asymmetry can be generated by precursor oscillations (see fig.~\ref{fig:overviewPlot}c(iii)). In this paper, we seek to understand how the parameters $\dot{\mu}$, $\da$ and $\thetaosc$ combine to affect the asymmetry of snap-through.

\paragraph{Summary of results with $\da=0$.} In ref.~\cite{qiong2024transient}, a buckled arch with perfectly symmetric boundary conditions (i.e.~$\da=0$) was considered. Some typical numerical results of this case are shown in fig.~\ref{fig:oscillationPlot}. It was observed that the size of the asymmetry measured during snap-through (via $\theta$) is proportional to the size of the precursor oscillations, $\thetaosc$: snap-through only amplifies a preexisting asymmetry caused by precursor oscillations. This is clear when running different numerical simulations at the same end-shortening rate $\dot{\mu}$, but introducing precursor oscillations of different amplitudes; when rescaling $\theta$ by $\thetaosc$, the curves collapse, as shown in the second row of figure \ref{fig:oscillationPlot}a) and b). Furthermore, fig.~\ref{fig:oscillationPlot}c) shows that as the rate of the unloading is decreased, the amplification of the asymmetry during snap-through transition increases, with a change in behaviour at a critical rate $\dot{\mu}_c$. 

\paragraph{Questions in the imperfect case  $\da\neq0$.} Given  the preceding summary of the arch with  perfectly symmetric boundary conditions, it is natural to ask what happens in the asymmetric case, $\da\neq0$. How does a small \emph{intrinsic} imperfection affect the transient symmetry during snap through? We will also consider how the presence of precursor oscillations interacts with an intrinsic asymmetry.

To answer these questions, we perform numerical simulations with a small value of $\da$. We shall first consider the case in which only imperfections are present, i.e.~no precursor oscillations, $\thetaosc=0$, before considering cases in which both $\da,\thetaosc>0$. Finally, in section \ref{sec:SimpleModel}, we will use a toy model for which theoretical solutions can be derived to explain the observations made. 

\subsection{Imperfection-driven transient asymmetry}

\begin{figure}[h!]
\includegraphics[width=\columnwidth]{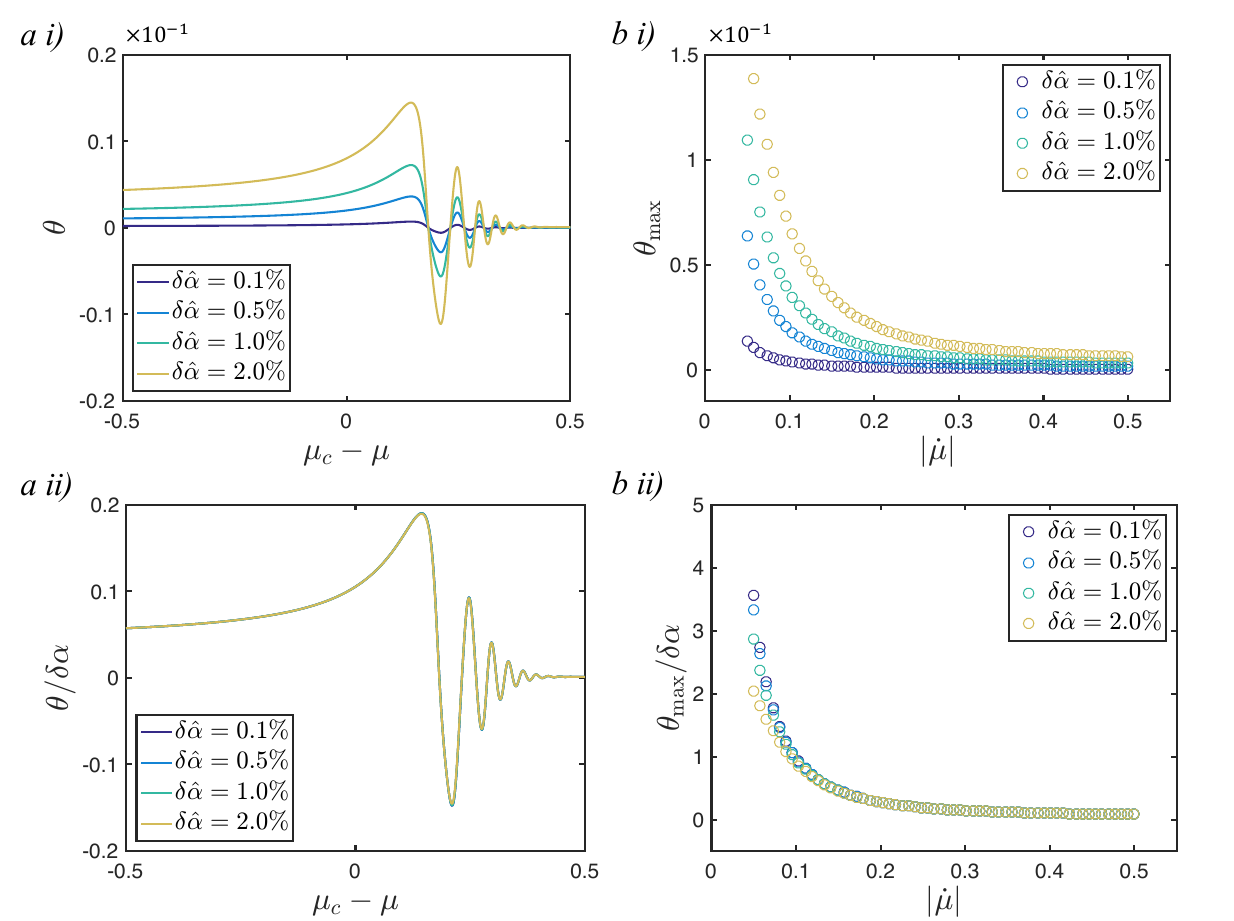}
\caption{Numerical simulations of the snap-through of an arch with intrinsic  imperfection $\da$ but no precursor oscillations i.e.~$\thetaosc=0$. (a)i) The evolution of the mid-point angle for different values of the imperfection $\delta\hat{\alpha}=\delta\alpha/\alpha$ but fixed $\dot{\mu}=0.25$ shows that larger imperfections in boundary conditions lead to greater asymmetry during snap-through. (a)ii) Rescaling $\theta$ by the imperfection size, $\delta\alpha$, all curves collapse onto one, suggesting that the midpoint growth is proportional to the size of the imperfection.  (b)i) The growth of the asymmetry depends on the unloading rate  $\dot{\mu}$, with smaller rates leading to larger asymmetry. (b)ii) Rescaling $\thetamax$ by the imperfection size, $\delta \alpha$, all curves collapse. A small divergence is observed at the smallest unloading rates, which we attribute to non-linearity.}
\label{fig:imperfectionPlot} 
\end{figure}

We begin by considering the case of asymmetry that is driven purely by imperfection, i.e.~we take $\thetaosc=0$. Figure \ref{fig:imperfectionPlot}a(i) shows how $\theta$ evolves as the end-shortening $\mu$ is changed at a relatively slow rate $|\dot{\mu}|=0.25$ for different values of the initial imperfection.  The asymmetry peaks during snap-through (i.e.~while $|\mu-\mu_c|$ is small) and the maximum value of $\theta$ increases with $\da$. Figure \ref{fig:imperfectionPlot}b shows how slower unloading rates lead to larger maximum angles $\thetamax$ with fixed $\da$.

Crucially, if we rescale $\theta$ by $\da$, all curves collapse (fig.~\ref{fig:imperfectionPlot}a(ii)), suggesting that the effect of intrinsic imperfections is also amplified during snap-through, similarly to precursor oscillations. We call the quantity 
$$G=\thetamax/\delta{\alpha}$$
the 'growth of the asymmetry'.  
Fig.~\ref{fig:imperfectionPlot}b(ii) shows that plotting the maximum growth of asymmetry (i.e.~rescaling $\thetamax$ by $\delta\alpha$) also collapses the data,  provided $\thetamax$ remains small. (There is a small deviation when $\thetamax$ becomes large, which we attribute to nonlinear effects.)

\subsection{Combining intrinsic imperfections with precursor oscillations }

Having seen that imperfections lead to growth of asymmetry and knowing from previous work \cite{qiong2024transient} that precursor oscillations have a similar effect,  we now explore how the combination of both precursor oscillations and imperfections can affect the system. When do imperfections dominate asymmetry, and when do precursor oscillations drive the growth of asymmetry?

\begin{figure}[h!]
\includegraphics[width=\columnwidth]{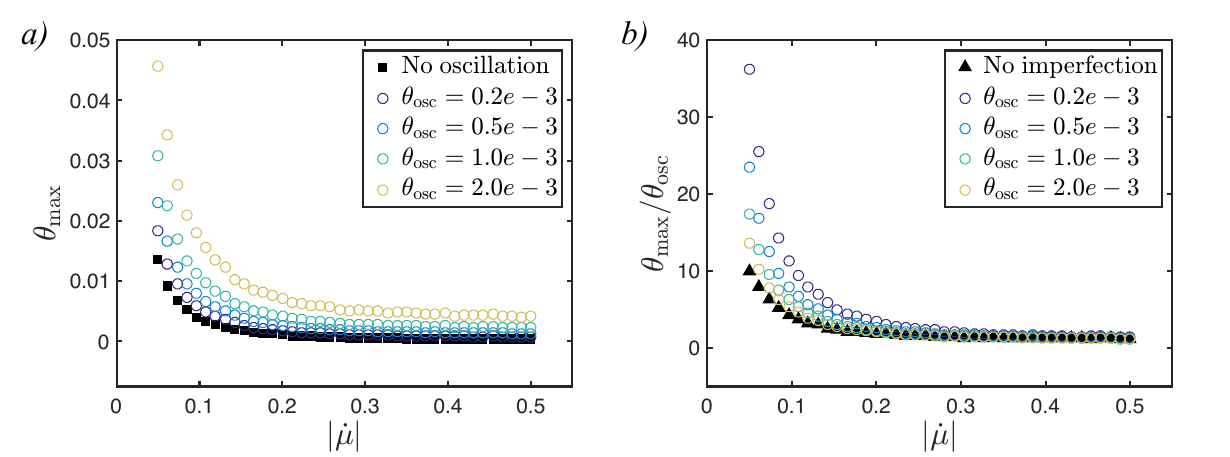}
\caption{Numerical simulations of the snap-through of an asymmetric arch with boundary angle imperfections $\delta \hat{\alpha}=0.001$ and asymmetric precursor oscillations. (a) Larger precursor oscillations lead to larger maximum midpoint angles. (b) Rescaling by the size of the oscillation we see that with relatively large precursor oscillations, our results converge towards those of \citet{qiong2024transient}. This suggests that, if the amplitude of the oscillation is sufficiently large, it dominates the physics and the imperfection does not matter.}
\label{fig:oscillationImpPlot} 
\end{figure}

In fig.~\ref{fig:oscillationImpPlot}a, we plot numerical results of the observed $\thetamax$ as a function of the rate $|\dot{\mu}|$ with different precursor oscillation size, $\thetaosc$, but fixed imperfection, $\da=0.001$. As expected, larger precursor oscillations lead to larger  asymmetry being observed. Following previous work \cite{qiong2024transient}, we again  define the amplification
$$\mathcal{A}=\thetamax/\thetaosc,$$ and plot (figure \ref{fig:oscillationImpPlot}b) this amplification as a function of loading rate. We observe that when the precursor oscillations are sufficiently large, $\thetaosc\gtrsim10^{-3}(=\da)$ in this case, the amplification $\mathcal{A}$ seems to be a function of $|\dot{\mu}|$ only --- the results collapse onto previous results for a completely symmetric system (black squares show data from ref.~\cite{qiong2024transient}). As might be expected, therefore, for large imperfections and small precursor oscillations, imperfections dominate the growth of asymmetry. Conversely, when precursor oscillations are sufficiently large, we see convergence to the imperfection-free system, meaning precursor oscillations dominate the behaviour. 

We have now seen that the symmetry of the arch as it snaps-through depends on three key parameters: the size of the precursor oscillations, $\thetaosc$, the size of the imperfection introduced in the clamps, $\da$, and the unloading rate, $|\dot{\mu}|$. To understand how these three parameters interact, we now develop a simplified model.

\section{Simple model\label{sec:SimpleModel}}

\begin{figure}[h!]
\includegraphics[width=0.8\columnwidth]{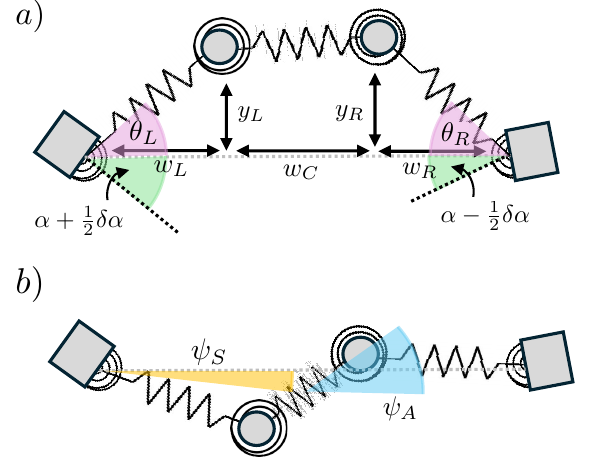}
\caption{Schematic diagram showing the double-mass von Mises truss used to understand the relative importance of an imperfection of size $\delta\alpha$ and precursor oscillations.}
\label{fig:schematicSM} 
\end{figure}

Solving analytically for the dynamic evolution of the arch is difficult: the problem is, at best, a partial differential equation in two variables (space and time) with a nonlinear end-shortening constraint. To gain insight into the behaviour of the system we instead use a simple model that was successfully used to gain understanding of the perfectly symmetric problem: a double mass von-Mises Truss \cite{qiong2024transient, zhang2016reconfiguration}. A schematic of the double mass von-Mises Truss is shown in figure \ref{fig:schematicSM}. The system is composed of two masses attached to two clamps and to each other by linear springs of stiffness $k$. The side springs have natural length $l_0$ while the central spring has natural length $l_{0c}$. The reference length of the system is $l=2l_0+l_{c}$ and we assume that the end distance can be changed by an amount $\delta l$.
To provide bending rigidity to the system, torsion springs of stiffness $B$ (and that favour a flat configuration) are connected about each mass. Finally, two more torsion springs are attached to the clamps, imposing a favoured angle with respect to the horizontal of $\alpha+ \half \delta \alpha$ and $\alpha- \half \delta \alpha$ at the left and right boundary, respectively. The latter springs mimic the clamped boundaries of the full arch problem and break both the left-right and up-down symmetries of the system.

As we shall show, when the clamp angles are small $\alpha, \delta \alpha \ll 1$ and given the right choice of parameters $L_{0C}=l_c/l_{0c}$ and $B$, the double mass von-Mises Truss reproduces the key behaviours of the full arch problem, but with the added benefit and simplicity of having only two degrees of freedom: the vertical positions of the two masses. To derive this behaviour, we use Lagrangian mechanics to derive the equations of motion (from the elastic and kinetic energies) and study the dynamic behaviour of the resulting system.

\subsection{Elastic Potential Energy}

In the deformed state of the von-Mises Truss we let $l_R$ and $l_L$ be the lengths of the left and right springs, respectively, while  $w_L$ and $w_R$ are the horizontal distances of the left mass and right mass from their respective clamps, $\Delta h$ is the difference in height between the two masses, $l_C$ the length of the central spring and $w_C$ its horizontal width. We also let $\theta_L$ and $\theta_R$ be the angles the side springs make with the horizontal, while $\theta_{LC}$ and $\theta_{RC}$ denote the angles between the side springs and the central one.  We also write $\Delta \theta$ for the angle the central spring makes with the horizontal.
Using elementary geometry, we then have that:
\begin{align}
    l_L&=\frac{w_L}{\cos\theta_L},\\
    l_R&=\frac{w_R}{\cos\theta_R},\\
    w_C&=l-\delta l-w_L-w_R,\\
    \Delta h=&w_L \tan\theta_L-w_R \tan\theta_R,\\
    l_C&=\sqrt{\Delta h^2+w_C^2},\\
    \Delta \theta&=\arctan(w_C,\Delta h),\\
    \theta_{LC}&=(\pi/2-\theta_{L})+(\pi/2-\Delta \theta),\\
    \theta_{RC}&=(\pi/2-\theta_{R})+(\pi/2+\Delta \theta).
\end{align} (Note that here $\arctan(x,y)$ is the two-argument arctangent; that is, for $x$, $y$ in the correct quadrant, $\arctan(x,y)=\arctan(y/x)$.)
We can use these relations to write the energy stored in the linear springs as:
\begin{align}
    U_{\mathrm{springs}}=\frac{k}{2}\left[(l_L-l_0)^2+ (l_R-l_0)^2+ (l_C-l_{0C})^2 \right]
\end{align}
while the bending energy stored in the torsion springs is
\begin{align}
\notag
    U_{B}=&\frac{B}{2}[(\theta_L-\alpha(1+\half \da))^2+(\theta_R-\alpha(1-\half \da))^2]\\
    &+\frac{B}{2}[(\theta_{LC}-\pi)^2+(\theta_{RC}-\pi)^2].
\end{align} 
The total energy of the system, given by $U=U_{\mathrm{springs}}+U_B$,  can be written as a function of the variables $w_L$, $w_R$, $\theta_L$, $\theta_R$ as well as the control parameter $\delta l$:
\begin{widetext}\begin{align}
\notag
 U=&\frac{1}{2} k \left[l_{0C}-\sqrt{\left(l- \delta l- w_L- w_R\right)^2+\left(w_L \tan \theta_L-w_R \tan \theta_R \right)^2}\right]^2
  +\frac{1}{2} k\left(\left(l_0-w_L \sec \theta_L\right)^2+\left(l_0-w_R \sec \theta _R\right){}^2\right)\\ 
    \notag
  &+\frac{1}{2} B \left[\arctan \left(l-\delta l-w_L-w_R,w_L \tan \theta_L-w_R \tan \theta_R \right)+\theta_L\right]^2\\
    \notag
    &+\frac{1}{2} B\left[\arctan \left(l-\delta l-w_L-w_R,w_L \tan \theta_L-w_R \tan \theta _R\right)-\theta _R\right]^2+\frac{1}{2} B \left[\left(\theta_L-\alpha(1+\half \da) \right)^2+\left(\theta_R -\alpha(1-\half \da)\right)^2\right].
   \label{eqn:totalenergy}
\end{align}
\end{widetext}

To make progress, we begin by using a suitable change of variable to emphasize any asymmetry. In particular, we introduce the symmetric and asymmetric parts of the angle and of the width, which we identify with the subscripts $(\cdot)_S$ and $(\cdot)_A$, respectively. In particular, we let
\begin{align}
    \theta_L&=\theta_S+\half \theta_A\\
    \theta_R&=\theta_S-\half \theta_A\\
    w_L&=l_0-\half(\delta l-w_S+w_A)\\
    w_R&=l_0-\half(\delta l-w_S-w_A).
\end{align}

\subsection{Energetic expansion for   $\alpha\ll1$}

When the arch profile is close to flat, i.e.~for  $\alpha\ll1$, all other angles, (i.e.~ $\theta_R$, $\theta_L$ and $\Delta \theta$) must also be small. Clearly, $\theta_R$, $\theta_L$ will be of the same size as $\alpha$. 
Conversely, $w_R$ and $w_L$ scale with $\alpha^2$. To see this, consider the unconstrained system in which the length $l$ is free, $\alpha$ is small, and the system can relax. In this case,  $\theta_i \sim \alpha$ for $i=L,R$. If we do not constrain the end-to-end displacement, the springs will be relaxed so that $l_i=l_0$ and the width changes according to $w_i \sim l_0 \cos \theta_i \sim l_0(1 - \half \alpha^2)$. We can thus rescale angles by $\alpha$ and changes in horizontal lengths by $\alpha^2$. This motivates introducing:
\begin{align}
\notag
    \theta_L&=\alpha (\psi_S-\half \psi_A)\\
    \notag
    \theta_R&=\alpha(\psi_S+\half \psi_A)\\
    \notag
    w_L&=l_0\bigl[1-\half \alpha^2 (\mu - W_S-W_A)\bigr]\\
    \notag
    w_R&=l_0\bigl[1-\half \alpha^2 (\mu - W_S+W_A)\bigr]\\
    \notag
    l_{0C}&=l_0 L_{0C}\\
    B&=\third \beta \alpha^2 k l_0^2
    \label{eqn:rescalings}
\end{align}
where $\mu=\delta l/(\alpha^2 l)$, $\psi_A=\theta_A/\alpha$, $\psi_S=\theta_S/\alpha$, $W_S=w_S/(l_0 \alpha^2)$, $W_A=w_A/(l_0 \alpha^2)$. We note, in particular, that the bifurcation parameter $$\mu=\frac{\delta l}{\alpha^2 l}$$ has emerged again, just as in the continuous case.

For the stretching and bending energy to be of the same magnitude, we need $B \sim \beta \alpha^2 k l_0^2$ (the factor of one-third being included to simplify later equations) \cite{qiong2024transient}. We then expand the energy to leading order (i.e.~at $O(\alpha^4)$) and find that:
\begin{align}
\label{eq:energyW}
192\tilde{U}&=\frac{24 W_S
   \left[(L_{0C}-4) \psi _A^2+4
   L_{0C} \left(\mu +2 \mu 
   L_{0C}+\psi
   _S^2\right)\right]}{L_{0C}}\\
   \notag
   &+48
   W_A^2+144
   W_S^2-96 W_A \psi _A \psi _S+ g_U,
\end{align} 
where $g_U$ is a function of $(\psi_S,\psi_A,\mu; \da)$. Equation \eqref{eq:energyW} is easily minimized with respect to variations in $W_S$ and $W_A$ by solving $\p{U}{W_A}=0$ and $\p{U}{W_S}=0$, leading to:
\begin{align}
\notag
    W_S&= -\frac{(L_{0C}-4) \psi _A^2+4 L_{0C}
   \left[\mu(1 +2 L_{0C})+\psi
   _S^2\right]}{12 L_{0C}},\\
\notag
   W_A &= \psi _A \psi _S,
\end{align}
and allowing us to write the energy only as a function of $\psi_A$ and $\psi_S$:

\begin{align}
\notag
\tilde{U}=&\half a_0 \da\, \psi_A- a_1 \psi_S+\half (a_2-a_3 \mu)\psi_S^2+\fourth a_4 \psi_S^4\\
&+\half (b_1 -b_2 \mu+b_3 \psi_S^2)\psi_A^2+\fourth b_4 \psi_A^4+U_0,
\label{eqn:simpleenergy}
\end{align}
where:
\begin{align}
    U_0&= \frac{1}{6}\left[2 \beta+(2+L_{0C})^2 \mu^2 \right],\\
    a_0&=\frac{\beta}{3}, \quad a_1= \frac{2}{3} \beta,\quad a_2= \frac{4}{3} \beta,\\
    a_3&= 2(2+L_{0C})/3,\quad a_4=\frac{2}{3},\\
    b_1&= \frac{2+L_{0C}(2+L_{0C})}{3 L_{0c}^2}\beta,\\
    b_2&= \frac{(2+L_{0C})^2}{6 L_{0C}},\quad     b_3=\frac{2+L_{0C}}{6 L_{0C}},\\
    b_4&=\frac{(2+L_{0C})^2}{24 L_{0C}^2}.
\end{align}
Note that the signs of terms in \eqref{eqn:simpleenergy} have been chosen so that all of the above constants are positive.

If we let $\da=0$, we return to the case studied by \citet{qiong2024transient}, who showed that the bifurcation structure of the continuous problem is quantitatively recovered by choosing $\beta=0.110$  and $L_{0C}=0.655$ (i.e.~with these values, the double von-Mises Truss becomes unstable to asymmetric perturbations at $\mu=1/4=\mu_c$ and the inverted equilibrium is lost in a saddle-node bifurcation at $\mu=0.247=\mu_\ast$). We therefore choose these values  hereafter.

  
  Having constructed an elastic energy that reproduces the desired static bifurcation structure, we now turn to the dynamics, which requires a consideration of the kinetic energy of the elements of the truss model.

\subsection{Kinetic energy}

The kinetic energy of the system consists of the translational and rotational energy of each of the two, identical, masses. This energy  is therefore easily written down by considering the motion (in polar coordinates) of the masses about the two edge elements; we find that
\begin{equation}
    T=\half m \left[(l_L \dot{\theta}_L)^2+\dot{l}_L^2+(l_R \dot{\theta}_R)^2+\dot{l}_R^2\right]
\end{equation}
where dots indicate derivatives with respect to time $t$. We use the rescalings from equation \eqref{eqn:rescalings} but must  also choose an appropriate rescaling of time. A natural choice, which allows a semi-quantitative comparison between the simple model and the full arch problem, is to rescale time by the period of oscillations in the system \cite{qiong2024transient}. Therefore, we let
\begin{equation}
    t=\frac{\tosc}{\alpha}\sqrt{\frac{m}{k}} \tilde{t},
\end{equation} 
where $ \tilde{t}$ is the dimensionless time, the scaling $\alpha^{-1}\sqrt{m/k}$ arises as a balance between bending and inertia terms, and 
\begin{equation}
    \tosc\approx9.477
    \label{eqn:toscDefn}
\end{equation} 
is a numerical prefactor as the dimensionless period of oscillations when $\mu=1/4$ \cite{qiong2024transient}.

Expanding the kinetic energy in small $\alpha$ we can write the rescaled kinetic energy, $\tilde{T}=T/(k l_0^2 \alpha^4)$, as
\begin{equation}
\tilde{T}=\frac{1}{{\tosc}}\left(\dot{\psi}_S^2+\fourth\dot{\psi}_A^2+H.O.T. \right).
\label{eqn:ke}
\end{equation} Then, writing the Lagrangian $L=T-U$ and minimizing with respect to $\psi_S$ and $\psi_A$ we find the equations describing the dynamical evolution of the system. Here, we write these equations under the assumption that the asymmetry is small (i.e.~$\psi_A\ll1$), only keeping the leading order term in $\psi_A$ and dropping all the tildes hereafter for convenience:
    \begin{align}
\label{eqn:dyn1B}
    \ddot{\psi}_S/{\tosc}^2&=\half a_1 - \half(a_2-a_3 \mu)\psi_S- \half a_4 \psi_S^3, \\
    \ddot{\psi}_A/{\tosc}^2&=a_0 \delta \hat{\alpha} -2 f(\psi_S;\mu) \psi_A, \label{eqn:dyn2B}
\end{align} where
\begin{equation}
f(\psi_S;\mu)=b_1 -b_2 \mu+b_3 \psi_S^2
\label{eqn:fDefn}.
\end{equation}

In the absence of imperfection, $\da=0$, the evolution of the asymmetry in \eqref{eqn:dyn2B} is entirely decided by the coefficient, $f(\psi;\mu)$ of $\psi_A$,  which in turn is governed by \eqref{eqn:dyn1B}. In particular, the sign of $f$ controls whether $\psi_A$ oscillates or grows exponentially. As shown by \citet{qiong2024transient} if the unloading is sufficiently fast, $f$ may remain positive throughout the motion  (because trajectories of the system  lag behind the equilibrium behaviour somewhat \cite{Liu2021}); in this case, therefore,  $\psi_A$ oscillates, but does not grow significantly during the motion. However, when the system is loaded slowly, $\psi_S$ remains close to its equilibrium value (no lag) and $f$  becomes negative for some portion of the motion. As a result, a small asymmetric perturbation, denoted  $\psi_A^0$ (equivalent to $\thetaosc$ in the full arch case), is amplified (exponentially) while $f<0$. Separating these two behaviours, there is a critical trajectory (observed when $|\dot{\mu}|_c\approx 0.96$) at which $f$ never becomes negative, but vanishes at an instant of time. This critical trajectory marks the sharp transition between weakly and highly amplified perturbations \cite{qiong2024transient}. In this `perfect' case, therefore the  symmetry, or otherwise, of the system is entirely governed by two parameters: the end-stretching rate, $|\dot{\mu}|$, and the size of the initial oscillation, $\psi_A^0$.

Once we allow for imperfections, $\da \neq 0$, in \eqref{eqn:dyn2B} we see that the transient growth of the asymmetric component $\psi_A$ is more complicated: the first term in the right hand side of eqn~\eqref{eqn:dyn2B} describes the importance of the intrinsic imperfection, while the second term describes the importance of the initial asymmetric shape perturbation, which may be the result of precursor oscillations. Clearly, the relative size of these two terms will be important, and so we expect the relative size of the imperfection, $\da$, and the initial size of the initial asymmetric shape perturbation, $\psi_A^0$, to play a role. As such, we introduce
\begin{equation}
    \gamma=\psi_A^0/\da.
    \label{eqn:GammaDefn}
\end{equation}



\begin{figure*}[t]
\includegraphics[width=\textwidth]{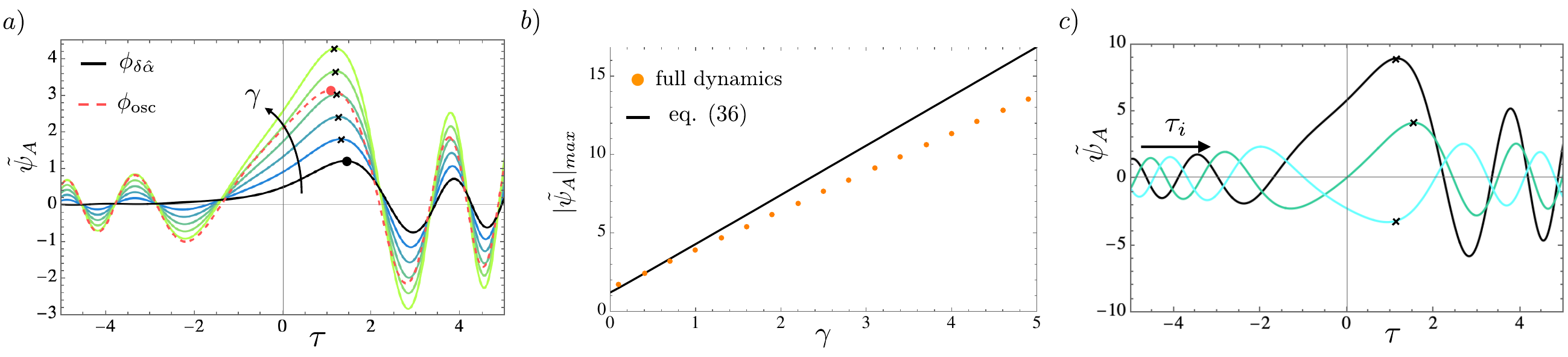}
\caption{Theoretical analysis of the asymmetric snap-through using the toy model. (a) The black line shows the solution of equation \eqref{eqn:phia} for $\phi_{\da}$ while the light-red dashed line shows the solution to equation \eqref{eqn:phiosc} for $\phiosc$. (The temporal maxima of $\phi_{\da}$ and $\phiosc$ are shown by the black and light-red dots, respectively.) The colored paths are obtained by solving equation \eqref{eqn:CriticalEqn}, derived by approximating the behaviour of the double-mass von-Mises truss at the critical loading rate $|\dot{\mu}|=|\dot{\mu}|_c$. The relative size of asymmetry introduced in the initial condition, $\gamma=\psi_A^0/\delta \alpha$, allows us to study the relative importance of imperfection versus precursor oscillations. (b) Plot of the maximum asymmetry $|\tilde{\psi_A}|_{\max}$ as a function of $\gamma$ as predicted by equation \eqref{eqn:CriticalEqn} compared to solution to the full problem, eqns \eqref{eqn:dyn1B} and \eqref{eqn:dyn2B} with $\da=0.001$. The relation is well approximated by a line with positive gradient, with precursor oscillations increasing the asymmetry. (c) When $\gamma>0$, changing the starting time of the evolution ($\tau_i$) changes the phase of the oscillation, and strongly influences the dynamics and thus the value of ${\tilde{\psi_A}}_{\max}$. In figures a) and b) the phase is chosen so that ${\tilde{\psi_A}}_{\max}$ is largest.}
\label{fig:theory} 
\end{figure*}

Our discussion suggests that three parameters determine the transient asymmetry of the system: the end-stretching rate, $|\dot{\mu}|$, and the size of the precursor oscillation, $\psi_A^0$, as in the perfect case, but also the size of the imperfection, $\da$. To study what happens in the transition from  imperfection-dominated to oscillation-dominated regimes, we need to study the nonlinear dynamics carefully. This problem is simplified by focussing on the behaviour at the critical end-stretching rate, $|\dot{\mu}|_c$, found in the perfect case, $\da=0$.

If we express both $\mu= -|\dot{\mu}|_c t$ and $\psi_{S}(t)$ explicitly as functions of time, then $f(\psi_S(t),\mu(t))=f(t)$ satisfies the condition $f(t^*)=f'(t^*)=0$ at some time $t^*$. We can expand our dynamic equation near this critical time by letting $t=t^*+\Delta t$. Keeping only the leading order terms in $\Delta t^2$, and rescaling equation \eqref{eqn:dyn2B} according to:
\begin{align}
    \Delta t = \frac{\tau}{(\tosc^2 \kappa)^{1/4}}, \quad \psi_A= \da \tilde{\psi}_A(\tau),
\end{align}
where $\kappa=f''(t^*)=0.819$,
we obtain the simple equation:
\begin{equation}
   \frac{\mathrm{d}^2 \tilde{\psi}_A}{\mathrm{d} \tau^2}=c-\tau^2 \tilde{\psi}_A,
   \label{eqn:CriticalEqn}
\end{equation}
where $c=a_0 \tosc/\sqrt{\kappa}\approx 0.385$.
When $\gamma=0$, the initial conditions at some initial time $\tau=-\tau_i$ require $c-\tau^2 \tilde{\psi}_A=0$, leading to $\tilde{\psi}_A(-\tau_i)=c/\tau_i^2$. Differentiating this equilibrium we find the other initial condition for the rate: $\tilde{\psi}_A'(-\tau_i)=-2 c/\tau_i^3$. To introduce oscillations, we simply alter the initial value of $\tilde{\psi}_A$ away from equilibrium, so that $\tilde{\psi}_A(-\tau_i)=c/\tau_i^2 + \gamma$. 

To solve the equation \eqref{eqn:CriticalEqn}, we exploit its linearity and decompose the solution as
\begin{equation} \tilde{\psi}_A=\phi_{\da}+\gamma \phiosc.
\label{eqn:decomp}
\end{equation}
where the first term captures the effects of the imperfection while the other that of precursor oscillations. The imperfections equation is:
\begin{equation}
   \frac{\mathrm{d}^2 \phi_{\da}}{\mathrm{d} \tau^2}=c-\tau^2 \phi_{\da},
   \label{eqn:phia}
\end{equation}
with initial conditions $\phi_{\da}(\tau_i)=c/\tau_i^2$ and $\phi'_{\da}(\tau_i)=-2c/\tau_i^3$. The growth of the asymmetry is simply a function of $\da$. This equation can be readily solved numerically. 

Conversely, the equation for $\phiosc$ reads:
\begin{equation}
   \frac{\mathrm{d}^2 \phiosc}{\mathrm{d} \tau^2}=-\tau^2 \phiosc,
   \label{eqn:phiosc}
\end{equation}
with initial conditions $\phiosc(-\tau_i)=1$ and $\dot{\phi}_{\mathrm{osc}}(-\tau_i)=0$. This equation has an analytical solution (given in the appendix \ref{sec:appA}), and describes the component of the evolution that is introduced by precursor oscillations.

The maximum absolute value of the two trajectories described by \eqref{eqn:phia} and 
\eqref{eqn:phiosc} are close to each other (at $\tau=1.41$ for $\phi_{\da}$ and at $\tau=1.17$ for $\phiosc$), meaning the maximum of $\tilde{\psi_A}$ is somewhere between these two values, as shown in figure \ref{fig:theory}. Thus, the growth of $\tilde{\psi_A}_{\max}$ as a function of $\gamma$ will be close to linear.


In figure \ref{fig:theory}(a) we show the evolution of $\tilde{\psi}_A$ as a function of time for various initial values of $\gamma$. When $\gamma=0$ there are no precursor oscillations and we observe only $\phi_{\da}$ (black line), while for larger values of $\gamma$, $\phiosc$ (shown in dashed light-red) introduces oscillations and increase the maximum asymmetry (marked by the black \emph{x} at the top of each coloured curve) during snap-through. 
In figure \ref{fig:theory}(b) we show how the maximum asymmetry $|\psi_A|_{\max}$ depends on $\gamma$ as predicted by the simplified dynamics of equation \eqref{eqn:CriticalEqn} (solid black line) and the full solution to the dynamics from equations \eqref{eqn:dyn1B} and \eqref{eqn:dyn2B} (orange dots). The plot suggests that there is no sharp transition between the imperfection-dominated and perturbation-dominated regime as one may have hoped. Instead, the size of the precursor oscillation influences the maximum asymmetry observed in a way that is close to linear in $\gamma$, at least for $\gamma\lesssim 1$.
Finally, note that the phase of the oscillation matters: in figure \ref{fig:theory}(c), we show that changing the starting point for the numerical solution of \eqref{eqn:CriticalEqn}, $\tau_i$, affects the phase of the oscillation, the dynamics, and thus $|\psi_A|_{\max}$. (In figure \ref{fig:theory}(a) and (b) we have chosen the phase to obtain the largest possible value of $|\psi_A|_{\max}$.)

Although our analysis has been carried out only at the critical rate $|\dot{\mu}|=|\dot{\mu}|_c$, a similar decomposition of $\psi_A$ as in equation \eqref{eqn:decomp} suggests that a near linear relationship between the amplitude of the precursor oscillations and the maximum $\theta$ is true also at different unloading rates. This behaviour indeed emerges for the full dynamic simulations of the simple model (as shown in figure \ref{fig:theory}b)) and, as we shall show in the conclusion, also in the full arch simulations. This means there is no clear transition from the imperfection-dominated to perturbation-dominated regime.

\section{Conclusions}

In this study, we analysed the problem of the snap-through of an elastic arch allowing for a combination of intrinsic boundary asymmetries (imperfections) and asymmetries introduced by the initial shape perturbations (precursor oscillations). We have then studied the question of when the resulting snap-through is asymmetric and, further, whether this asymmetry is dominated by imperfections or precursor oscillations. The answer to this apparently innocuous question is complex: using a toy model, we showed that it  depends  not only on the rate at which the control parameter $\mu$ is changed (as shown in previous work), but also on the relative size of any precursor oscillations to the magnitude of intrinsic imperfection, $\gamma$, defined in \eqref{eqn:GammaDefn}. Taking the equivalent form for the full arch problem, we let $\gamma=\thetaosc/\da$ and therefore expect that the maximum asymmetry, $\thetamax=\da\,G(\gamma;|\dot{\mu}|)$.

\begin{figure}[h!]
\includegraphics[width=0.9\columnwidth]{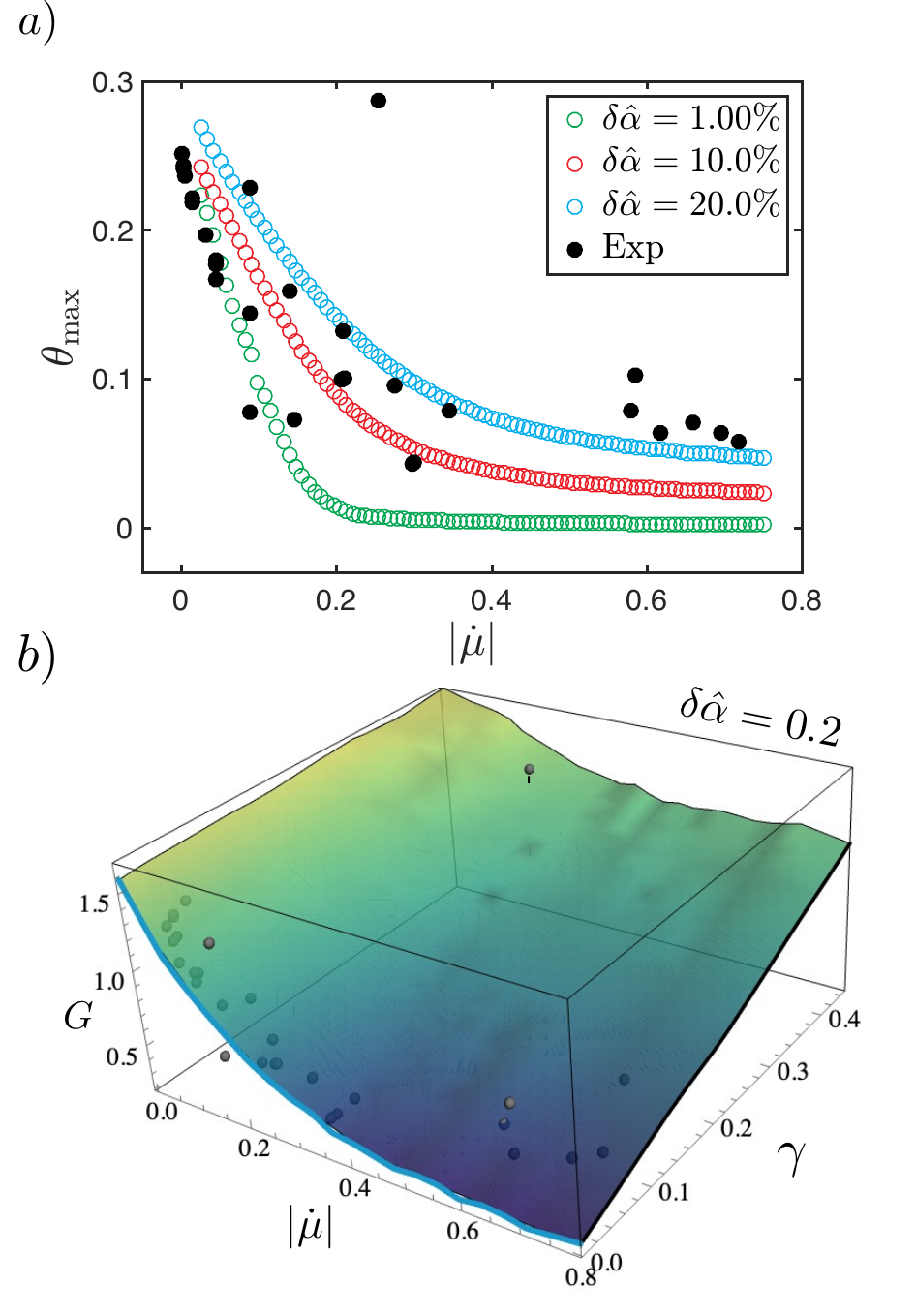}
\caption{
Comparison of previously-published experimental results \cite{qiong2024transient} with full numerical results incorporating the effect of intrinsic asymmetry. (a) Experimental results ($\bullet$)  and numerical results with different levels of asymmetry $\da$ ($\circ$ --- see legend) but no precursor oscillations (i.e.~$\thetaosc=\gamma=0$) show that experiments are generally consistent with the expected relative asymmetry $\da\approx0.2$, but also that anomalous results remain. (b) Combining with measured values of the precursor oscillation in experiments, and assuming $\da=0.2$, we calculate the relevant value of $\gamma=\thetaosc/\da$ and see that numerical results for the maximum angle produce a 3D surface for the predicted growth of asymmetry that is close to that observed experimentally. } 
\label{fig:expPlot} 
\end{figure}

To illustrate this complexity, we conclude by reconsidering the experimental results of \citet{qiong2024transient}. In fig.~\ref{fig:expPlot}a we plot the maximum observed asymmetry, $\thetamax$, as a function of loading rate.  This shows broad dispersion, but is generally consistent with numerical results if precursor oscillations are ignored (i.e.~$\gamma=0$) and we take $\da=0.2$ (broadly consistent with experiments where an angle of $\alpha=\pi/12\approx 15^\circ$ was used with a precision of $\pm2^\circ$). Nevertheless, when viewed as a two-dimensional slice at $\gamma=0$ this completely misses some experimental results, which therefore appear to be anomalous. Taking $\gamma=0$ ignores the more complicated picture that we have uncovered here using both numerical simulations and a toy model. Instead, plotting the three dimensional surface $G(\gamma;|\dot{\mu}|)$ with fixed $\da=0.2$ and using the experimentally measured value of $\thetaosc$ to determine $\gamma$ gives a different picture (see fig.~\ref{fig:expPlot}b):  the experimental points that are apparently anomalous actually correspond to experimental runs in which the initial precursor oscillation was larger than usual. (The size of the precursor oscillations, $\thetaosc$, is generally determined by the size of the acceleration required to generate a particular value of $|\dot{\mu}|$; hence $\thetaosc$ and $\gamma$ are generally larger for larger $|\dot{\mu}|$ in fig.~\ref{fig:expPlot}b. However, there are also experimental runs at more moderate $|\dot{\mu}|$ for which the precursor oscillations are anomalously large. For this reason it is important to measure $\thetaosc$ directly in experiments.)

This work shows that the combination of loading rate and precursor oscillation amplitude can be combined either to mitigate or enhance asymmetries caused by imperfections in elastic snap-through. This may give a new strategy for controlling the dynamic shapes adopted by snapping objects \cite{huang2024exploiting}. In particular, since it has proven challenging to control the magnitude of the precursor oscillations in experiments, we suggest that choosing the size of the imperfection to be large in comparison to  the typical size of the precursor oscillations may be the best way to control the asymmetry of the resulting dynamics. This approach also allows one to choose the left or right asymmetric mode during snap-through repeatably, allowing for control of the jumping direction.

\vspace{2mm}
\begin{acknowledgments} 
This work was partially supported by the UK Engineering and Physical Sciences Research Council via Grant No.~EP/ W016249/1 (A.G.~and D.V.). Q.W., Y.W., and S.T.~acknowledge support from the Defense Advanced Research Projects Agency DARPA SHRIMP HR001119C0042. W.H.~acknowledges  start-up funding from Newcastle University, UK. Y.W. acknowledges  support from A*STAR, Singapore. M.L.~acknowledges support via start-up funding from the University of Birmingham.  For the purpose of Open Access, the author has applied a CC BY public copyright licence to any Author Accepted Manuscript version arising from this submission.
\end{acknowledgments}

\appendix

\label{fig:newScale} 
\newpage
\section{Solutions to equation \eqref{eqn:phiosc}}
\label{sec:appA}
The analytical solution to the problem 
\begin{equation}
    \frac{\mathrm{d}^2 \phiosc}{\mathrm{d} \tau^2}=-\tau^2 \phiosc
\end{equation}
with initial conditions $\phiosc(-\tau_i)=1$ and $\dot{\phi}_{\mathrm{osc}}(-\tau_i)=0$ is given by
\begin{align}
    \phiosc=A D_{-\frac{1}{2}}\bigl[(1+\mi) \tau \bigr]+B
   D_{-\frac{1}{2}}\bigl[(-1+\mi) \tau \bigr]
\end{align}
where $\mi=\sqrt{-1}$, $D_{\nu}(x)$ is the Parabolic cylinder function \cite{Abramowitz1964,Olver2010} and $A$ and $B$ are given by:
\begin{widetext}
\begin{align}
    A&=\frac{(1+\mi) \tau_i D_{-\frac{1}{2}}\bigl[(1-\mi)\tau_i\bigr]-2 \mi D_{\frac{1}{2}}\bigl[(1-\mi)\tau_i\bigr]}{2 D_{-\frac{1}{2}}\bigl[(1-\mi)\tau_i\bigr] D_{\frac{1}{2}}\bigl[(-1-\mi) \tau_i\bigr]+D_{-\frac{1}{2}}\bigl[(-1-\mi) \tau_i\bigr]
   \left((2+2 \mi) \tau_i
   D_{-\frac{1}{2}}\bigl[(1-\mi) \tau_i\bigr]-2 \mi
   D_{\frac{1}{2}}\bigl[(1-\mi) \tau_i\bigr]\right)}\\
   B&=\frac{(1+\mi) \tau_i D_{-\frac{1}{2}}\bigl[(-1-\mi)\tau_i\bigr]+2 D_{\frac{1}{2}}\bigl[(-1-\mi)\tau_i\bigr]}{2 D_{-\frac{1}{2}}\bigl[(1-\mi)
   \tau_i\bigr] D_{\frac{1}{2}}\bigl[(-1-\mi) \tau_i\bigr]+D_{-\frac{1}{2}}\bigl[(-1-\mi) \tau_i\bigr]   \left((2+2 \mi) \tau_i
   D_{-\frac{1}{2}}\bigl[(1-\mi) \tau_i\bigr]-2 \mi
   D_{\frac{1}{2}}\bigl[(1-\mi) \tau_i\bigr].\right)}
\end{align}
\end{widetext}

\section{The importance of phase}

\begin{figure}[h!]
\includegraphics[width=\columnwidth]{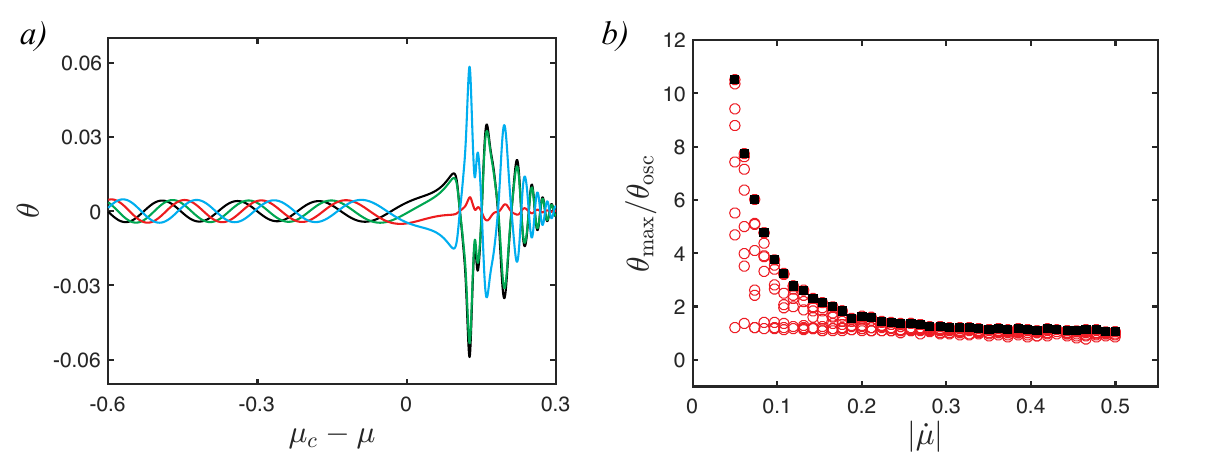}
\caption{(a) Simulations of snap-though of an arch with no imperfection $\da=0$ and at the rate $|\dot{\mu}| = 0.1$. The simulations are initiated at slightly different values of $\mu$, leading to different phases of the oscillation and ultimately different maximum asymmetries being achieved during the transient dynamics. b) for each value of $\dot{\mu}$ we run several simulations varying the phase of the oscillation and calculate the amplification $\theta_{\max}/\thetaosc$ (red circles); we then select the largest amplification obtained to calculate the envelope (black squares).  }
\label{fig:phasePlots}
\end{figure}

The asymmetry achieved by the arch during snap-through in the presence of precursor oscillations depends significantly on the phase of the oscillation as the system passes through the snap-through point --- see fig.~\ref{fig:phasePlots}a. Oscillations that are leading to increase in asymmetry as the system undergoes snap-though undergo large amplification while those  that are in the decreasing phase at snap-through are only weakly amplified. When evaluating the asymmetry obtained at a given rate, we consider the largest possible asymmetry that can be achieved. To obtain this maximal asymmetry, we run several simulations with precursor oscillations of the same amplitude (same $\thetaosc$), but different phases and select the largest value to create the envelope shown by the black squares in fig.~\ref{fig:phasePlots}b.

\bibliography{paper.bib}

\end{document}